\documentclass[12pt]{article}
\usepackage{graphicx}
\begin{document}
\begin{center}
{\large\bf Modified Gravitational Theory and Galaxy Rotation Curves}
\vskip 0.3 true in {\large J. W. Moffat}
\vskip 0.3 true in {\it The Perimeter Institute for Theoretical Physics,
Waterloo, Ontario, N2J 2W9, Canada} \vskip 0.3 true in and \vskip 0.3 true
in {\it Department of Physics, University of Toronto, Toronto, Ontario M5S
1A7, Canada} \end{center}
\begin{abstract}%
The nonsymmetric gravitational theory predicts an acceleration law
that modifies the Newtonian law of attraction between particles.
For weak fields a fit to the flat rotation curves of galaxies is
obtained in terms of the mass (mass-to-light ratio $M/L$) of
galaxies. The fits assume that the galaxies are not dominated by
exotic dark matter. The equations of motion for test particles
reduce for weak gravitational fields to the GR equations of motion
and the predictions for the solar system and the binary pulsar PSR
1913+16 agree with the observations. The gravitational lensing of
clusters of galaxies can be explained without exotic dark matter.
\end{abstract}
\vskip 0.2 true in e-mail: jmoffat@perimeterinstitute.ca


\section{Introduction}

A gravitational theory explanation of the acceleration of the
expansion of the universe~\cite{Perlmutter,Riess,Spergel} and the
observed flat rotation curves of galaxies was
proposed~\cite{Moffat}, based on the nonsymmetric gravitational
theory (NGT)~\cite{Moffat2,Moffat3,Moffat4}. Since no dark matter
has been detected so far, it seems imperative to seek a possible
modified gravitational theory that could explain the now large
amount of data on galaxy rotation curves. The same holds true for
the need to explain the acceleration of the expansion of the
universe without having to invoke a cosmological constant, because
of the serious problems related to this constant~\cite{Weinberg}.

In the following, we summarize the derivation of the motion of
test particles in NGT. We consider the derivation of test particle
motion from the NGT conservation laws. The motion of a particle in
a static, spherically symmetric gravitational field is derived,
yielding the modified Newtonian law of motion for weak
gravitational fields. Two parameters $\sqrt{M_0}$ and $r_0$ occur
in the generalized Newtonian acceleration law. The parameter
$\sqrt{M_0}$ is modelled for a bound system by a dependence on the
mean orbital radius of a test particle, and the range parameter
$r_0$ is determined for galaxies and clusters of galaxies from the
acceleration $cH_0$ where $H_0$ is the measured Hubble constant. A
fit to both low surface brightness and high surface brightness
galaxies is achieved in terms of the total galaxy mass $M$ (or
$M/L$) without exotic dark matter. A satisfactory fit is achieved
to the rotational velocity data generic to the elliptical galaxy
NGC 3379. Fits to the data of the two spheroidal dwarf galaxies
Fornax and Draco and the globular cluster $\omega$ Cenauri are
also obtained. The predicted light bending and lensing can lead to
agreement with galaxy cluster lensing observations.

The modelled values of the parameter $\sqrt{M_0}$ for the solar
system and Earth, lead to agreement with solar system
observations, terrestrial gravitational experiments and the binary
pulsar PSR 1913+16 observations.

\section{The Field Equations}

The nonsymmetric fundamental tensor $g_{\mu\nu}$ is defined
by~\cite{Moffat2,Moffat3,Moffat4}:
\begin{equation}
g_{\mu\nu}=g_{(\mu\nu)}+g_{[\mu\nu]},
\end{equation}
where
\begin{equation}
g_{(\mu\nu)}={1\over 2}(g_{\mu\nu}+g_{\nu\mu}),\quad g_{[\mu\nu]}=
{1\over 2}(g_{\mu\nu}-g_{\nu\mu}).
\end{equation}
The nonsymmetric connection $\Gamma^\lambda_{\mu\nu}$ is
decomposed as
\begin{equation}
\Gamma^\lambda_{\mu\nu}=\Gamma^\lambda_{(\mu\nu)}
+\Gamma^\lambda_{[\mu\nu]}.
\end{equation}
The contravariant tensor $g^{\mu\nu}$ is defined in terms of the
equation
\begin{equation}
\label{inverse}
g^{\mu\nu}g_{\sigma\nu}=g^{\nu\mu}g_{\nu\sigma}={\delta^\mu}_\sigma.
\end{equation}

The NGT action is given by
\begin{equation}
S_{\hbox{ngt}}=S+S_M,
\end{equation}
where
\begin{equation}
\label{NGTaction} S=\frac{1}{16\pi G}\int d^4x[{\bf
g}^{\mu\nu}R^*_{\mu\nu}(W)-2\Lambda\sqrt{-g} -{1\over 4}\mu^2{\bf
g}^{\mu\nu}g_{[\nu\mu]}],
\end{equation}
and $S_M$ is the matter action satisfying the relation
\begin{equation}
\frac{1}{\sqrt{-g}}\biggl(\frac{\delta S_M}{\delta
g^{\mu\nu}}\biggr)=-\frac{1}{2}T_{\mu\nu}.
\end{equation}
Here, we have chosen units $c=1$, ${\bf
g}^{\mu\nu}=\sqrt{-g}g^{\mu\nu}$, $g=\hbox{Det}(g_{\mu\nu})$,
$\Lambda$ is the cosmological constant, $\mu$ is a mass associated
with the skew field $g_{[\mu\nu]}$. Moreover, $T_{\mu\nu}$ is the
nonsymmetric energy-momentum tensor and $R^*_{\mu\nu}(W)$ is the
tensor
\begin{equation}
R^*_{\mu\nu}(W)=R_{\mu\nu}(W)-\frac{1}{6}W_\mu W_\nu,
\end{equation}
where $R_{\mu\nu}(W)$ is the NGT contracted curvature tensor
\begin{equation}
R_{\mu\nu}(W)=W^\beta_{\mu\nu,\beta} - {1\over
2}(W^\beta_{\mu\beta,\nu}+W^\beta_{\nu\beta,\mu}) -
W^\beta_{\alpha\nu}W^\alpha_{\mu\beta} +
W^\beta_{\alpha\beta}W^\alpha_{\mu\nu},
\end{equation}
defined in terms of the unconstrained nonsymmetric connection:
\begin{equation}
\label{Wequation}
W^\lambda_{\mu\nu}=\Gamma^\lambda_{\mu\nu}-{2\over
3}{\delta^\lambda}_\mu W_\nu,
\end{equation}
where
\begin{equation}
W_\mu={1\over 2}(W^\lambda_{\mu\lambda}-W^\lambda_{\lambda\mu}).
\end{equation}
Eq.(\ref{Wequation}) leads to the result
\begin{equation}
\Gamma_\mu=\Gamma^\lambda_{[\mu\lambda]}=0.
\end{equation}
The contracted tensor $R_{\mu\nu}(W)$ can be written as
\begin{equation}
R_{\mu\nu}(W)=R_{\mu\nu}(\Gamma)+\frac{2}{3}W_{[\mu,\nu]},
\end{equation}
where
\begin{equation}
R_{\mu\nu}(\Gamma ) = \Gamma^\beta_{\mu\nu,\beta} -{1\over 2}
\left(\Gamma^\beta_{(\mu\beta),\nu} +
\Gamma^\beta_{(\nu\beta),\mu}\right) - \Gamma^\beta_{\alpha\nu}
\Gamma^\alpha_{\mu\beta} +
\Gamma^\beta_{(\alpha\beta)}\Gamma^\alpha_{\mu\nu}.
\end{equation}

The gravitational constant $G$ in the action $S$ is defined in
terms of the ``bare'' gravitational constant $G_0$:
\begin{equation}
\label{gravirenorm} G=G_0Z,
\end{equation}
where $G_0=6.673\times 10^{-8}\,{\rm g}^{-1}\,{\rm cm}^3\,{\rm
s}^{-2}$ is Newton's constant and $Z$ depends on the strength of
the coupling of $g_{[\mu\nu]}$ to matter. Thus, $Z=1$ when
$g_{[\mu\nu]}$ is zero and NGT reduces to Einstein's GR.

A variation of the action $S_{\hbox{ngt}}$ yields the field
equations in the presence of matter sources
\begin{equation}
\label{Gequation} G^*_{\mu\nu} (W)+\Lambda g_{\mu\nu}+S_{\mu\nu}
=8\pi GT_{\mu\nu},
\end{equation}
\begin{equation}
\label{divg} {{\bf g}^{[\mu\nu]}}_{,\nu}=-\frac{1}{2}{\bf
g}^{(\mu\alpha)}W_\alpha,
\end{equation}
\begin{equation}
{{\bf g}^{\mu\nu}}_{,\sigma}+{\bf g}^{\rho\nu}W^\mu_{\rho\sigma}
+{\bf g}^{\mu\rho} W^\nu_{\sigma\rho}-{\bf
g}^{\mu\nu}W^\rho_{\sigma\rho}
$$ $$
+{2\over 3}\delta^\nu_\sigma{\bf g}^{\mu\rho}W^\beta_{[\rho\beta]}
+{1\over 6}({\bf g}^{(\mu\beta)}W_\beta\delta^\nu_\sigma -{\bf
g}^{(\nu\beta)}W_\beta\delta^\mu_\sigma)=0.
\end{equation}
Here, we have $G^*_{\mu\nu}(W)=R^*_{\mu\nu}(W) - {1\over 2}
g_{\mu\nu}{\cal R}^*(W)$, where ${\cal
R}^*(W)=g^{\mu\nu}R^*_{\mu\nu}(W)$, and
\begin{equation}
S_{\mu\nu}=\frac{1}{4}\mu^2(g_{[\mu\nu]} +{1\over
2}g_{\mu\nu}g^{[\sigma\rho]}
g_{[\rho\sigma]}+g^{[\sigma\rho]}g_{\mu\sigma}g_{\rho\nu}).
\end{equation}

The vacuum field equations in the absence of matter sources are
given by
\begin{equation}
\label{vacuumeqs}
 R^*_{\mu\nu}(W)=\Lambda
g_{\mu\nu}-(S_{\mu\nu}-\frac{1}{2}g_{\mu\nu}{\cal S}),
\end{equation}
where ${\cal S}=g^{\mu\nu}S_{\mu\nu}$. For the case of a static,
spherically symmetric field with $\mu=0$ and $g_{[0i]}=0$
(i=1,2,3), the gravitational field is described by the Wyman
solution given in Appendix A~\cite{Wyman}. In this case the vector
$W_\mu=0$~\cite{Moffat,Moffat3,Moffat4} and the field equations
(\ref{vacuumeqs}) in empty space $T_{\mu\nu}=0$ become
\begin{equation}
\label{emptyspace} R_{\mu\nu}(\Gamma)=\Lambda
g_{\mu\nu}-(S_{\mu\nu}-\frac{1}{2}g_{\mu\nu}{\cal S}).
\end{equation}
The time independent components of these field equations are given
in Appendix A.

\section{The Equations of Motion of a Particle}

In NGT there are two choices for the motion of a
particle~\cite{Legare}, for there exist two connections, the
Christoffel connection given by
\begin{equation}
\left\{{\lambda\atop \mu\nu}\right\}={1\over 2}s^{(\lambda\rho)}
\left(g_{(\mu\rho),\nu}+g_{(\rho\nu),\mu}-g_{(\mu\nu),\rho}\right),
\end{equation}
where
\begin{equation}
s^{(\nu\alpha)}g_{(\mu\alpha)}={\delta^\nu}_\mu,
\end{equation}
and the nonsymmetric connection $\Gamma^\lambda_{\mu\nu}$. If we
define the parallel transport of a vector $V^\mu$ by
\begin{equation} D_\mu V^\lambda=\partial_\mu
V^\lambda+\Gamma^\lambda_{\rho\mu}V^\rho, \end{equation} then we
obtain the equation of motion \begin{equation}
\label{pathequation} u^\nu D_\nu u^\lambda\equiv
\frac{du^\lambda}{d\tau}+\Gamma^\lambda_{\mu\nu}u^\mu u^\nu=0,
\end{equation} where $\tau$ is the proper time along the path
followed by the particle and $u^\lambda=dx^\lambda/d\tau$ is the
4-velocity of the particle. This defines the path equation which
is not a path of extremal length.

We can derive the second equation of motion by using the
Lagrangian defined by
\begin{equation}
L=g_{(\mu\nu)}u^\mu u^\nu.
\end{equation}
We have
\begin{equation}
\label{extremal}
\frac{1}{2}\biggl(\frac{d}{d\tau}\frac{\partial L}{\partial
u^\alpha} -\frac{\partial L}{\partial
x^\alpha}\biggr)=g_{(\mu\nu)}u^\mu u^\nu
\biggl(\left\{{\beta\atop
\mu\nu}\right\}-\Gamma^\beta_{\mu\nu}\biggr). \end{equation} We
see that when the right-hand side vanishes, the Euler-Lagrangian
equations are satisfied and we get the extremal geodesic equation
\begin{equation} \label{geodesic}
\frac{du^\mu}{d\tau}+\left\{{\mu\atop\alpha\beta}\right\}u^\alpha
u^\beta=0. \end{equation}

The conservation laws in NGT can be written~\cite{Legare2}:
\begin{equation}
\label{conservation}
\partial_\rho{{\tilde{\bf T}}^\rho}_\lambda
-\frac{1}{2}\partial_\lambda g_{\mu\nu}{\bf T}^{\mu\nu}=0,
\end{equation}
where
\begin{equation}
{{\tilde{\bf T}_\lambda}}^\rho
=\frac{1}{2}(g_{\mu\lambda}{\bf T}^{\mu\rho}
+g_{\lambda\mu}{\bf T}^{\rho\mu}).
\end{equation}

Let us assume a monopole test particle
\begin{equation}
\int d^3x{\bf T}^{\mu\nu}\not= 0,
\end{equation}
\begin{equation}
\int d^3x(x^\alpha-X^\alpha){\bf T}^{\mu\nu}=0,
\end{equation}
and similarly for higher moments. The integration is carried out over a
hypersurface of constant $t$, following the procedure of
Papapetrou~\cite{Papapetrou}, and $X^\alpha$ is the position of the
monopole. It can be shown that (\ref{conservation}) leads to
\begin{equation}
\label{eqmotion}
\frac{d}{dt}\biggl(\frac{dX^\beta}{dt}\int d^3x{\bf
t}^{00}\biggr) + \left\{{\beta\atop
\mu\nu}\right\}\frac{dX^\mu}{dt}\frac{dX^\nu}{dt}\int d^3x{\bf
t}^{00}
$$ $$
=\frac{1}{2}s^{(\lambda\beta)}\biggl(\partial_\lambda
g_{\mu\nu}\int d^3x{\bf T}^{\mu\nu}-\partial_\lambda
g_{(\mu\nu)}\frac{dX^\mu}{dt} \frac{dX^\nu}{dt}\int
d^3x{\bf t}^{00}\biggr),
\end{equation}
where ${\bf t}^{\mu\nu}=s^{(\lambda\nu)}{{\tilde{\bf
T}_\lambda}}^\mu$. If we assume that ${\bf T}^{[\mu\nu]}=0$,
then we get
\begin{equation}
\label{geodesicequation}
\frac{d}{d\tau}\biggl(m\frac{dX^\beta}{d\tau}\biggr)+m\left\{{\beta\atop
\mu\nu}\right\}\frac{dX^\mu}{d\tau}\frac{dX^\nu}{d\tau}=0,
\end{equation}
where
\begin{equation} m=\frac{d\tau}{dt}\int
d^3x{\bf t}^{00}.
\end{equation}
This leads us to the geodesic equation (\ref{geodesic}). However,
we cannot in general assume that $T^{[\mu\nu]}$ is zero, for it is
associated with the intrinsic spin of matter or to the skew field
$g_{[\mu\nu]}$ itself, and we must include a coupling to the spin
and skew field on the right-hand side of (\ref{geodesicequation}).
We shall find that in the weak field approximation the path
equation (\ref{pathequation}) and the geodesic equation
(\ref{geodesic}) become approximately the same, including a
necessary coupling to a skew symmetric source.

\section{Particle Motion in a Static Spherically Symmetric
Gravitational Field}

It was shown in ref.~\cite{Legare} that for large values of $r$
in the static, spherically symmetric solution of the NGT field
equations, {\it the geodesic equation and the path equation
yield similar physical results}. Since we shall be concerned
with the dynamics of galaxies, we shall treat the case of the
geodesic equation coupled to a skew symmetric source. We have
\begin{equation}
\label{geodesic2}
\frac{du^\beta}{d\tau}+
\left\{{\beta\atop \mu\nu}\right\}u^\mu u^\nu
=s^{(\beta\alpha)}f_{[\alpha\mu]}u^\mu,
\end{equation}
where
\begin{equation}
f_{[\alpha\mu]}=\lambda\partial_{[\alpha}
\biggl(\frac{\epsilon^{\eta\sigma\nu\lambda}}{\sqrt{-g}}
H_{[\sigma\nu\lambda]}g_{(\mu]\eta)}\biggr).
\end{equation}
Here, the skew tensor $\epsilon^{\mu\nu\lambda\eta}$ is the
Levi-Civita tensor density and $H_{[\mu\nu\lambda]}$ is given by
\begin{equation}
H_{[\mu\nu\lambda]}=\frac{1}{3}(\partial_\lambda
g_{[\mu\nu]} +\partial_\mu g_{[\nu\lambda]}+\partial_\nu
g_{[\lambda\mu]}),
\end{equation}
and $\lambda$ is a coupling
constant with the dimension of length that couples the skew
field to the test particle.

In the static, spherically symmetric field of NGT, the tensor
$H_{[\mu\nu\lambda]}$ has only one non-vanishing component
\begin{equation}
H_{[\theta\phi
r]}=\frac{1}{3}\partial_rg_{[\theta\phi]}=f'\sin\theta
\end{equation}
and we get
\begin{equation}
f_{[r0]}=\lambda\frac{d}{dr}\biggl(\frac{\gamma
f^\prime}{\sqrt{\alpha\gamma(r^4+f^2)}}\biggr).
\end{equation}

The equations of motion for a test particle are given by
\begin{equation}
\label{rmotion}
\frac{d^2r}{d\tau^2}+\frac{\alpha'}{2\alpha}\biggl(\frac{dr}{d\tau}\biggr)^2
-\frac{r}{\alpha}\biggl(\frac{d\theta}{d
\tau}\biggr)^2-r\biggl(\frac{\sin^2\theta}{\alpha}\biggr)\biggl(\frac{d\phi}{d\tau}\biggr)^2
+\frac{\gamma'}{2\alpha}\biggl(\frac{dt}{d\tau}\biggr)^2
$$ $$
+\frac{d}{dr}\biggl(\frac{\lambda\gamma
f'}{\sqrt{\alpha\gamma(r^4+f^2)}}\biggr)\biggl(\frac{dt}{d\tau}\biggr)=0,
\end{equation}
\begin{equation}
\label{tequation}
\frac{d^2t}{d\tau^2}+\frac{\gamma'}{\gamma}\biggl(\frac{dt}{d\tau}\biggr)
\biggl(\frac{dr}{d\tau}\biggr)+\frac{1}{\gamma}\frac{d}{dr}\biggl(\frac{\lambda\gamma
f'}{\sqrt{\alpha\gamma(r^4+f^2)}}\biggr)=0,
\end{equation}
\begin{equation}
\frac{d^2\theta}{d\tau^2}+\frac{2}{r}\biggl(\frac{d\theta}{d\tau}
\biggr)\biggl(\frac{dr}{d\tau}\biggr)-\sin\theta\cos\theta\biggl(\frac{d\phi}{d\tau}\biggr)^2
=0,
\end{equation}
\begin{equation}
\label{phiequation}
\frac{d^2\phi}{d\tau^2}+\frac{2}{r}\biggl(\frac{d\phi}{d\tau}\biggr)\biggl(\frac{dr}{d\tau}\biggr)
+2\cot\theta\biggl(\frac{d\phi}{d\tau}\biggr)\biggl(\frac{d\theta}{d\tau}\biggr)=0.
\end{equation}

The orbit of the test particle can be shown to lie in a plane and
by an appropriate choice of axes, we can make $\theta=\pi/2$.
Integrating Eq.(\ref{phiequation}) gives
\begin{equation}
\label{angular}
r^2\frac{d\phi}{d\tau}=J,
\end{equation}
where $J$ is the conserved orbital angular momentum.
Integration of Eq.(\ref{tequation}) gives
\begin{equation}
\label{dtequation}
\frac{dt}{d\tau}=-\frac{1}{\gamma}\biggl[\frac{\lambda\gamma
f'}{\sqrt{\alpha\gamma(r^4+f^2)}}+E\biggr],
\end{equation}
where $E$ is the constant energy per unit mass.

By substituting (\ref{dtequation}) into (\ref{rmotion}) and
using (\ref{angular}), we obtain
\begin{equation}
\label{reducedrmotion}
\frac{d^2r}{d\tau^2}+\frac{\alpha'}{2\alpha}\biggl(\frac{dr}{d\tau}\biggr)^2
-\frac{J^2}{\alpha
r^3}+\frac{\gamma'}{2\alpha\gamma^2}\biggl[\frac{\lambda\gamma
f'}{\sqrt{\alpha\gamma(r^4+f^2)}}+E\biggr]^2
=\frac{1}{\gamma}\frac{d}{dr}\biggl[\frac{\lambda\gamma
f'}{\sqrt{\alpha\gamma(r^4+f^2)}}+E\biggr].
\end{equation}

Let us now make the approximations that $\lambda f'/r^2\ll 1$,
$f/r^2\ll 1$ and the slow motion approximation $dr/dt\ll 1$.
Then, for material particles we set $E=1$
and (\ref{reducedrmotion}) becomes
\begin{equation}
\label{Newton} \frac{d^2r}{dt^2}-\frac{J_N^2}{r^3}+\frac{GM}{r^2}
=\lambda\frac{d}{dr}\biggl(\frac{f'}{r^2}\biggr),
\end{equation}
where $J_N$ is the Newtonian orbital angular momentum.

\section{Linear Weak Field Approximation}

We expand $g_{\mu\nu}$ about a Ricci-flat GR background
\begin{equation}
g_{\mu\nu}=g^{GR}_{(\mu\nu)}+h_{\mu\nu}+O(h^2),
\end{equation}
where $g^{GR}_{(\mu\nu)}$ is the GR background metric. The skew
field $h_{[\mu\nu]}$ obeys the linearized equation of motion in
the GR background geometry
\begin{equation}
\label{linearizedeq}
\nabla^\sigma
F_{\mu\nu\sigma}+4h^{[\sigma\beta]}B_{\beta\mu\sigma\nu}+\mu^2h_{[\mu\nu]}=0,
\end{equation}
where
\begin{equation}
F_{[\mu\nu\lambda]}=\partial_\lambda
h_{[\mu\nu]} +\partial_\mu h_{[\nu\lambda]}+\partial_\nu
h_{[\lambda\mu]}
\end{equation}
and $\nabla^\lambda$ and $B_{\beta\mu\sigma\nu}$
denote the background GR covariant derivative and curvature
tensor, respectively. Moreover,
\begin{equation}
W_\mu=-2\nabla^\lambda h_{[\mu\lambda]}.
\end{equation}
For flat Minkowski spacetime, Eq.(\ref{linearizedeq}) reduces
exactly to the massive Kalb-Ramond-Proca equation~\cite{Ramond},
which is free of ghost pole instabilities with a positive
Hamiltonian bounded from below.

For the static, spherically symmetric spacetime, the linearized equation of
motion (\ref{linearizedeq}) on a Schwarzschild background takes
the form
\begin{equation}
\biggl(1-\frac{2GM}{r}\biggr)f''-\frac{2}{r}\biggl(1-\frac{3GM}{r}\biggr)f'
-\biggl(\mu^2+\frac{8GM}{r^3}\biggr)f=0,
\end{equation}
where we assume that $\gamma(r)\sim 1/\alpha(r)\sim 1-2GM/r$ and
$\beta(r)=r^2$. The solution to this equation in leading order
is~\cite{Moffat3,Clayton2,Cornish}
\begin{equation}
f(r)=\frac{sG^2M^2}{3}\frac{\exp(-\mu r)}{(\mu r)^{\mu
GM}}\biggl(1+\mu r+ \frac{GM}{r}\biggl[2+\mu r\exp(2\mu
r)Ei(1,2\mu r)(\mu r-1)\biggr]\biggr),
\end{equation}
where $Ei$
is the exponential integral function
\begin{equation}
Ei(n,x)=\int_1^{\infty}dt\frac{\exp(-xt)}{t^n}.
\end{equation}
The constant $sG^2M^2/3$ is fixed from the exact Wyman solution
with $A=0$ by taking the limit $\mu\rightarrow 0$.

For large $r$ we obtain
\begin{equation}
f(r)=\frac{1}{3}\frac{sG^2M^2\exp(-\mu r)(1+\mu r)}{(\mu r)^{\mu
M}},
\end{equation}
and for $\mu GM\ll 1$, we have $(\mu r)^{\mu GM}\sim 1$, giving
\begin{equation}
f(r)=\frac{1}{3}sG^2M^2\exp(-\mu r)(1+\mu r).
\end{equation}
This is a solution to the equation
\begin{equation}
\label{fequation}
 f''(r)-\frac{2}{r}f'(r)-\mu^2 f(r)=0.
\end{equation}
If we consider the expansion about a GR Schwarzschild background,
then (\ref{fequation}) is valid for $\mu^2\gg 8GM/r^3$.

To the order of weak field approximation, we obtain from
Eq.(\ref{Newton}):
\begin{equation}
\label{Yukawa} \frac{d^2r}{dt^2}-\frac{
J_N^2}{r^3}=-\frac{GM}{r^2}+\frac{\sigma\exp(-\mu r)}{r^2}(1+\mu
r),
\end{equation}
where the constant $\sigma$ is given by
\begin{equation}
\label{sigma} \sigma=\frac{\lambda sG^2M^2\mu^2}{3}.
\end{equation}
Here, $\lambda$ and $s$ denote the coupling strengths of the test
particle and the skew $g_{[23]}$ field, respectively. In Eq.
(\ref{Yukawa}), we have required that the additional NGT
acceleration on the right-hand side is a repulsive force. This is
in keeping with the weak field approximation result obtained
in~\cite{Moffat2,Moffat3}, which corresponds to a skew force
produced by a massive axial vector spin $1^+$ boson exchange.

\section{Orbital Equation of Motion}

Let us write the line element as
\begin{equation}
ds^2=\gamma dt^2-\alpha dr^2-r^2(d\theta^2+\sin^2\theta d\phi^2).
\end{equation}
We set $\theta=\pi/2$ and divide the resulting expression by
$d\tau^2$ and use Eqs.(\ref{angular}) and (\ref{dtequation}) to
obtain
\begin{equation}
\label{energyconserved}
\biggl(\frac{dr}{d\tau}\biggr)^2+\frac{J^2}{\alpha
r^2}-\frac{1}{\alpha\gamma}\bigg[\frac{\lambda\gamma f'}
{\sqrt{\alpha\gamma(r^4+f^2)}}+E\biggr]^2=-\frac{E}{\alpha}.
\end{equation}
We have $ds^2=Ed\tau^2$, so that $ds/d\tau$ is a constant. For
material particles $E>0$ and for massless photons $E=0$.

Let us set $u=1/r$ and by using (\ref{angular}), we have
$dr/d\tau=-Jdu/d\phi$. Substituting this into
(\ref{energyconserved}), we obtain
\begin{equation}
\label{neworbital}
\biggl(\frac{du}{d\phi}\biggr)^2=
\frac{1}{\alpha\gamma J^2}\biggl[E+\frac{\lambda\gamma
f'}{\sqrt{\alpha\gamma(r^4+f^2)}}\biggr]^2-\frac{1}{\alpha
r^2}-\frac{E}{\alpha J^2}.
\end{equation}
By substituting $dr/d\phi=-(1/u^2)du/d\phi$ into
(\ref{neworbital}) and using the approximation $\gamma\sim
1/\alpha\sim 1-2GM/r$ and $f/r^2\ll 1$ and $\lambda f'/r^2\ll 1$,
we get after some manipulation
\begin{equation}
\label{finalorbital}
\frac{d^2u}{d\phi^2}+u=\frac{EGM}{J^2}-\frac{E\lambda
sG^2M^2}{3r_0^2J^2}\exp\biggl(-\frac{1}{r_0u}\biggr)\biggl(1+\frac{1}{r_0u}\biggr)
+3GMu^2,
\end{equation}
where $r_0=1/\mu$.

For material test particles $E=1$ and we obtain
\begin{equation}
\label{materialorbit}
\frac{d^2u}{d\phi^2}+u=\frac{GM}{J^2}+3GMu^2-\frac{K}{J^2}\exp\biggl(-\frac{1}{r_0u}\biggr)
\biggl(1+\frac{1}{r_0u}\biggr),
\end{equation}
where $K=\lambda sG^2M^2/3r_0^2$. On the other hand, for massless
photons $ds^2=0$ and $E=0$ and (\ref{finalorbital}) gives
\begin{equation}
\label{photons}
\frac{d^2u}{d\phi^2}+u=3GMu^2.
\end{equation}

\section{Galaxy Rotational Velocity Curves}

A possible explanation of the galactic rotational velocity curves
problem has been obtained in NGT~\cite{Sokolov}. From the radial
acceleration derived from (\ref{Yukawa}) experienced by a test
particle in a static, spherically symmetric gravitational field
due to a point source, we obtain
\begin{equation}
\label{accelerationlaw}
a(r)=-\frac{G_{\infty}M}{r^2}+\sigma\frac{\exp(-r/r_0)}{r^2}
\biggl(1+\frac{r}{r_0}\biggr),
\end{equation}
Moreover, $G\equiv G_{\infty}$ is defined to be the gravitational
constant at infinity
\begin{equation}
\label{renormG}
G_{\infty}=G_0\biggl(1+\sqrt{\frac{M_0}{M}}\biggr),
\end{equation}
where $G_0$ is Newton's ``bare'' gravitational constant. This
conforms with our definition of $G$ in Eq.(\ref{gravirenorm}),
which requires that $G$ be renormalized in order to guarantee that
(\ref{accelerationlaw}) reduces to the Newtonian acceleration
\begin{equation}
\label{Newtonianacceleration}
a_{\rm Newton}=-\frac{G_0M}{r^2}
\end{equation}
at small distances $r\ll r_0$. The constant $\sigma$ is given by
\begin{equation}
\label{sigma2}
\sigma=\frac{\lambda s G_0^2M^2}{3c^2r_0^2}.
\end{equation}

The integration constant $s$ in (\ref{sigma2}), occurring in the
static, spherically symmetric solution (see, Appendix A), is
dimensionless and can be modelled as
\begin{equation}
s=gM^a,
\end{equation}
where $M$ is the total mass of the particle source, $g$ is a
coupling constant and $a$ is a dimensionless constant. We choose
$a=-3/2$ and $\lambda gG_0^2/3c^2r_0^2=G_0\sqrt{M_0}$ where $M_0$
is a parameter. The choice of $a=-3/2$ yields for a galaxy
dynamics the Tully-Fisher law~\cite{Tully}.

We obtain the acceleration on a point particle
\begin{equation}
\label{accelerationlaw2}
a(r)=-\frac{G_{\infty}M}{r^2}+G_0\sqrt{MM_0}\frac{\exp(-r/r_0)}{r^2}
\biggl(1+\frac{r}{r_0}\biggr).
\end{equation}
By using (\ref{renormG}), we can write the NGT acceleration in the
form
\begin{equation}
\label{NGTaccelerate}
a(r)=-\frac{G_0M}{r^2}\biggl\{1+\sqrt{\frac{M_0}{M}}\biggl[1-\exp(-r/r_0)
\biggl(1+\frac{r}{r_0}\biggr)\biggr]\biggr\}.
\end{equation}

We conclude that the gravitational constant can be different at
small and large distance scales depending on the size of the
parameter $\sqrt{M_0}$. We have two parameters: the parameter
$\sqrt{M_0}$ and the distance range $r_0$. We assume that
$G_{\infty}$ scales for constant $M$ with increasing strength as
$\sqrt{M_0}$, while for fixed values of $M_0$ we have
$G_{\infty}\rightarrow G_0$ as the total mass of the source
$M\rightarrow\infty$. Let us model the parameter
$\alpha=\sqrt{M_0}$ as a function of the mean orbital radius of a
test particle
\begin{equation}
\label{alphaeq} \alpha\equiv \sqrt{M_0}=k\langle r_{\rm
orb}\rangle^n,
\end{equation}
where $k$ and $n$ are constants. We shall choose the exponent to
be $n=3/2$. We shall apply (\ref{NGTaccelerate}) to explain the
flatness of rotation curves of galaxies, as well as the
approximate Tully-Fisher law~\cite{Tully}, $G_0M\sim v^4$, where
$v$ is the rotational velocity of a galaxy and $M$ is the galaxy
mass
\begin{equation}
M=M_*+M_{HI}+M_{DB}+M_f.
\end{equation}
Here, $M_*,M_{HI}$, $M_{DB}$ and $M _f$ denote the visible mass,
the mass of neutral hydrogen, possible dark baryon mass and gas,
and the mass from the skew field energy density, respectively.
In~\cite{Moffat}, we obtained from the modified Friedmann
equations
\begin{equation}
\Omega=\Omega_b+\Omega_m+\Omega_f,
\end{equation} where
$\Omega_b$, $\Omega_m$ and $\Omega_f$ denote the fractional
values of baryons, $g_{[\mu\nu]}$ field matter density and
smooth $g^0_{[\mu\nu]}$ field density background, respectively,
obtained from the expansion
\begin{equation}
g_{[\mu\nu]}=g^0_{[\mu\nu]}+\delta g_{[\mu\nu]},
\end{equation}
where the smooth background $g^0_{[\mu\nu]}$ and the fluctuations
$\delta g_{[\mu\nu]}$ describe the ``dark energy'' and ``dark
matter'' components, respectively. The NGT cosmological field
equations must be solved to yield the observational results
$\Omega_b\sim 0.05$, $\Omega_m\sim 0.3$, and $\Omega\sim 0.7$ in
order to be consistent with WMAP and supernovae
data~\cite{Perlmutter,Riess,Spergel}. The quantities $\Omega_m$
and $\Omega_f$ in NGT replace the dark matter and dark energy. The
mass $M_f$ obtained from the skew field density $\rho_m$ is
expected to contribute to the total mass $M$ of galaxies.

The rotational velocity of a star $v$ is given by
\begin{equation}
\label{rotvelocity}
v=\sqrt{\frac{G_0M}{r}}\biggl\{1+\sqrt{\frac{M_0}{M}}\biggl[1-\exp(-r/r_0)
\biggl(1+\frac{r}{r_0}\biggr)\biggr]\biggr\}^{1/2}.
\end{equation}

Let us postulate that the parameters $M_0$ and $r_0$ give the
magnitude of the constant acceleration
\begin{equation}
\label{specialacceleration}
a_0=\frac{G_0M_0}{r^2_0}.
\end{equation}
We assume that for galaxies and clusters of galaxies this
acceleration is determined by
\begin{equation}
\label{Hubbleacceleration} a_0=cH_0.
\end{equation}
Here, $H_0$ is the
current measured Hubble constant $H_0=100\, h\, {\rm km}\,
s^{-1}\,{\rm Mpc}^{-1}$ where $h=(0.71\pm 0.07)$~\cite{pdata}.
This gives
\begin{equation}
a_0=6.90\times 10^{-8}\,{\rm cm}\,
s^{-2}.
\end{equation} A good fit to low surface brightness and
high surface brightness galaxy data is obtained with the
parameters
\begin{equation}
\label{parameters} M_0=9.60\times
10^{11}M_{\odot},\quad r_0=13.92\,{\rm kpc}=4.30\times
10^{22}\,{\rm cm}
\end{equation}
and $M$ (or the mass-to-light ratio $M/L$). Substituting
$M_0=9.60\times 10^{11}M_{\odot}$ into (\ref{specialacceleration})
yields the $r_0$ in (\ref{parameters}). Thus, we fit the galaxy
rotation curve data with one parameter $M_0$ and the total galaxy
mass $M$. Since we are using an equation of motion for point
particle sources, we are unable to fit the cores of galaxies. A
possible model for the galaxy cores is to assume for a radius
$r\ll r_c$, where $r_c$ is the core radius, a rotation curve of an
isothermal sphere in the ideal case where we can consider a
massless disk embedded in it. Then, for $ r\ll r_c$:
\begin{equation}
v(r)\sim \biggl(\frac{4\pi G_0\rho_c}{3}\biggr)^{1/2}r,
\end{equation}
where $\rho_c$ is the core density. For $r\gg r_c$, the rotational
velocity curve will be described by the NGT model
(\ref{rotvelocity}). Further investigation of this issue will
require solving the field equations of NGT for a core mass density
profile and will be the subject of future research.

We can now fix the constant $k$ in Eq.(\ref{alphaeq}) by using the
relation
\begin{equation}
k=\frac{\alpha_g}{\langle r_{\rm orb}\rangle_g^{3/2}},
\end{equation}
where $\alpha_g$ and $\langle r_{\rm orb}\rangle_g$ denote the
values of these quantities for galaxies, respectively. For the
mean value $\langle r_{\rm orb}\rangle_g$ we choose
\begin{equation}
\langle r_{\rm orb}\rangle_g=200\,{\rm kpc}
\end{equation}
and using the value of $M_0$ in (\ref{parameters}) we obtain
\begin{equation}
k=2.02\times 10^{-13}\,{\rm g}^{1/2}\,{\rm cm}^{-1}.
\end{equation}

The fits to the galaxy rotation curves $v$ in km/s versus the
galaxy radius $r$ in kps are shown in Fig. 1. The data are
obtained from ref.~\cite{McGaugh}.

In Fig. 2, fits to two dwarf galaxies (dSph) are shown. We assume
that the relation between the velocity dispersion $\sigma$ and the
rotational velocity $v$ takes the simple form in e.g. an
isothermal sphere model for which $v\sim \sqrt{2}\sigma$. The
error bars on the data~\cite{Lokas} for the velocity dispersions
are large, and in the case of Draco, due to the small radial range
$0.1\,{\rm kpc} < r < 0.6\,{\rm kpc}$, the Newtonian curve for
\begin{equation}
v=\sqrt{\frac{G_0M}{r}},
\end{equation}
cannot be distinguished within the errors from the NGT prediction.
However, it is noted that the NGT prediction for $v$ appears to
flatten out as $r$ increases. For Draco
$M/L=28.93+50.30(9.58)(M_{\odot}/L_{\odot})$, whereas for Fornax
$M/L=1.79+0.72(0.40)(M_{\odot}/L_{\odot})$. There is also an
expected large error in the distance estimates to the dSph.
Another serious potential source of error is that it is assumed
that dSph galaxies are in dynamical equilibrium. The two studied
here are members of the Local Group and exist in the gravitational
field of a larger galaxy, the Milky Way. Thus, the tidal
interactions with the larger galaxy are expected to affect the
dynamics of dSph galaxies and the interpretations of velocity
dispersions~\cite{Mateo}. These issues and others for dSph
galaxies are critically considered in the context of dark matter
models by Kormendy and Freeman~\cite{Kormendy}.

We have also included a fit to the elliptical galaxy NGC 3379. The
elliptical galaxy NGC 3379 has been the source of controversy
recently~\cite{Romanowsky}. The velocities of elliptical galaxies
are randomly distributed in the galaxy. However, the gravitational
potential that would be experienced by a test particle star or
planetary nebula in circular rotation about the center of the
galaxy can be extracted from the line-of-sight velocity dispersion
profiles. The data for $R/R_{\rm eff}> 0.5$ refer to planetary
nebula.

We use the mean values of the extracted rotational velocities for
NGC 3379, obtained by Romanowsky et al.~\cite{Romanowsky} and find
that the NGT predicted rotational velocities agree well with their
data. According to Romanowsky et al. there appears to be a dearth
of dark matter in the elliptical galaxy which needs to be
explained by dark matter models and N-body cosmological
simulations. For Milgrom's MOND~\cite{Milgrom,Bekenstein} it is
argued by Sanders and Milgrom~\cite{Sanders} that NGC 3379 is
marginally within the MOND regime with an acceleration $a\sim
(a_0)_{\rm Milgrom}\sim 1.2\times 10^{-8}\,{\rm cm}\,{\rm
s}^{-2}$, so MOND should not apply to the elliptical galaxies. As
we see from the fit to the data, the NGT results agree well with
the data. Romanowsky et al. also give data for the two elliptical
galaxies NGC 821 and NGC 4494, but the intrinsic circular
velocities associated with the line-of-sight velocity dispersion
profile data are not given by the authors, although the trends of
the data are similar to NGC 3379.

A fit to the data for the globular cluster $\omega$ Centauri is
shown in Fig. 3. The data is from McLaughlin and
Meylan~\cite{Meylan}. We use the velocity dispersion data and
assume that the data is close to the rotational velocity curves
associated with the velocity dispersion $\sigma_p$ i.e. the
isothermal sphere model relation $v\sim \sqrt{2}\sigma_p$ holds.
The fit to the data reveals that the predicted rotational velocity
cannot be distinguished from the Newtonian-Kepler circular
velocity curve within the orbital radius of the data. The authors
conclude that there appears to be no room for Milgrom's MOND or
dark matter, whereas the NGT results agree well with the data.

In Fig.4, we display a 3-dimensional plot of $v$ versus the range
of distance $0.1\,{\rm kpc} < r < 10\,{\rm kpc}$ and the range of
total galaxy mass $M$ used in the fitting of rotational velocity
data. The red surface shows the Newtonian values of the rotational
velocity $v$, while the black surface displays the NGT prediction
for $v$.

Table 1, displays the values of the total mass $M$ used to fit
the galaxies and the mass-to-light ratios $M/L$ estimated from
the data in references given in~\cite{McGaugh}.

There are now about 100 galaxies with available rotational
velocity data~\cite{McGaugh}. However, some of these galaxies are
not well described by spherically symmetric halos or have some
other disturbing physical feature, so we are unable to obtain fits
to these data.

In Milgrom's phenomenological model~\cite{Milgrom,Bekenstein} we
have
\begin{equation}
\label{Milgromv}
v^4=G_0M(a_0)_{\rm Milgrom},
\end{equation}
where
\begin{equation}
(a_0)_{\rm Milgrom}=1.2\times 10^{-8}\,{\rm cm}\, s^{-2}.
\end{equation}
We see that (\ref{Milgromv}) predicts that the rotational velocity
is constant out to an infinite range and the rotational velocity
does not depend on a distance scale, but on the magnitude of the
acceleration $(a_0)_{\rm Milgrom}$. In contrast, the NGT
acceleration formula does depend on the radius $r$ and the
distance scale $r_0$ which for galaxies is fixed by the formula
(\ref{Hubbleacceleration}).

\section{Local and Solar System Observations}

We obtain from Eq.(\ref{materialorbit}) the orbit equation
\begin{equation}
\label{particleorbit}
\frac{d^2u}{d\phi^2}+u=\frac{GM}{c^2
J^2}-\frac{K}{J^2}\exp(-r/r_0)\biggl[1
+\biggl(\frac{r}{r_0}\biggr)\biggr]+\frac{3GM}{c^2}u^2,
\end{equation}
where now $K=\lambda sG^2M^2/3c^4r_0^2$.
Using the large $r$ weak field approximation, and the expansion
\begin{equation}
\exp(-r/r_0)=
1-\frac{r}{r_0}+\frac{1}{2}\biggl(\frac{r}{r_0}\biggr)^2+...
\end{equation}
we obtain the orbit equation for $r<r_0$:
\begin{equation}
\label{orbitperihelion} \frac{d^2u}{d\phi^2}+u=\frac{GM}{c^2J_N^2}
-\frac{K}{J_N^2}+3\frac{GM}{c^2}u^2.
\end{equation}
We can write this as
\begin{equation}
\label{perihelioneq}
\frac{d^2u}{d\phi^2}+u=N+3\frac{GM}{c^2}u^2,
\end{equation}
where
\begin{equation}
N=\frac{GM}{c^2J_N^2}-\frac{K}{J_N^2}.
\end{equation}

We can solve Eq.(\ref{perihelioneq}) by perturbation theory and
find for the perihelion advance of a planetary orbit
\begin{equation}
\label{perihelionformula} \Delta\omega=\frac{6\pi}{c^2p}
(GM_{\odot}-c^2K_{\odot}),
\end{equation}
where $J_N=(GM_{\odot}p/c^2)^{1/2}$, $p=a(1-e^2)$ and $a$ and $e$
denote the semimajor axis and the eccentricity of the planetary
orbit, respectively.

We choose for the mean orbital radius of the solar planetary
system
\begin{equation}
\langle r_{\rm orb}\rangle_{\odot}=1\,{\rm a.u.}=1.49\times
10^{13}\,{\rm cm},
\end{equation}
which yields using (\ref{alphaeq}) and $k=2\times 10^{-13}\,{\rm
g}^{1/2}\,{\rm cm}^{-1}$:
\begin{equation}
\alpha_{\odot}\equiv(\sqrt{M_0})_{\odot}=1.2\times 10^7\,{\rm
g}^{1/2}
\end{equation}
and
\begin{equation}
\biggl(\sqrt{\frac{M_0}{M}}\biggr)_{\odot}=2.6\times 10^{-10}.
\end{equation}
This result gives
\begin{equation}
\label{deltaG}
\frac{\Delta G}{G_0}\sim 10^{-10}.
\end{equation}
and $G=G_0$ within the experimental errors for the measurement of
Newton's constant $G_0$.

We choose for the solar system $K_{\odot}\ll 1.5\,{\rm km}$ and
use $G=G_{\infty}=G_0$ to obtain from (\ref{perihelionformula}) a
perihelion advance of Mercury in agreement with GR.

For terrestrial experiments and orbits of satellites, we choose
the mean orbital radius equal to the earth-moon distance, $\langle
r_{\rm orb}\rangle_{\oplus}=3.57\times 10^{10}\,{\rm cm}$, which
gives
\begin{equation}
\biggl(\sqrt{\frac{M_0}{M}}\biggr)_{\oplus}=1.7\times 10^{-11}.
\end{equation}
This also yields $G=G_0$ within the experimental errors.

For the binary pulsar PSR 1913+16 the formula
(\ref{perihelionformula}) can be adapted to the periastron shift
of a binary system. Combining this with the NGT gravitational wave
radiation formula, which will approximate closely the GR formula,
we can obtain agreement with the observations for the binary
pulsar.  We choose the mean orbital radius equal to the projected
semi-major axis of the binary, $\langle r_{\rm orb}\rangle_{\rm
N}=7\times 10^{10}\,{\rm cm}$, giving $(\sqrt{M_0/M})_N=5\times
10^{-14}$. Thus, $G=G_0$ within the experimental errors and
agreement with the binary pulsar data for the periastron shift is
obtained for $K_N\ll 4.2\,{\rm km}$.

For a massless photon $E=0$ and we have
\begin{equation}
\label{lightbending}
\frac{d^2u}{d\phi^2}+u=3\frac{GM}{c^2}u^2.
\end{equation}
For the solar system using (\ref{deltaG}) this gives the light
deflection:
\begin{equation}
\Delta_{\odot}=\frac{4G_0M_{\odot}}{c^2R_{\odot}}
\end{equation}
in agreement with GR.

\section{Galaxy Clusters and Lensing}

We can assess the existence of dark matter of galaxies and
clusters of galaxies in two independent ways: from the dynamical
behavior of test particles through the study of extended rotation
curves of galaxies, and from the deflection and focusing of
electromagnetic radiation, e.g., gravitational lensing of clusters
of galaxies. The light deflection by gravitational fields is a
relativistic effect, so the second approach provides a way to test
the relativistic effects of gravitation at the extra-galactic
level. It has been shown that for conformal tensor-scalar gravity
theories the bending of light is either the same or can even be
weaker than predicted by GR~\cite{Bekenstein,Bekenstein2}. To
remedy this problem, Bekenstein~\cite{Bekenstein} has recently
formulated a relativistic description of Milgrom's MOND model,
including a time-like vector field as well as two scalar fields
within a GR metric scenario. However, the time-like vector field
violates local Lorentz invariance and requires preferred frames of
reference.

The bending angle of a light ray as it passes near a massive
system along an approximately straight path is given to lowest
order in $v^2/c^2$ by
\begin{equation}
\label{lensingformula} \theta=\frac{2}{c^2}\int\vert
a^{\perp}\vert dz,
\end{equation}
where $\perp$ denotes the perpendicular component to the ray's
direction, and dz is the element of length along the ray and $a$
denotes the acceleration. The best evidence in favor of dark
matter lensing is the observed luminous arcs seen in the central
regions of rich galaxy clusters~\cite{Blanford}. The cluster
velocity dispersion predicted by the observed arcs is consistent
within errors with the observed velocity dispersion of the cluster
galaxies. This points to a consistency between the virial mass and
the lensing mass, which favors the existence of dark matter.

From (\ref{lightbending}), we obtain the
light deflection
\begin{equation}
\Delta=\frac{4GM}{c^2R}=\frac{4G_0{\overline M}}{c^2R},
\end{equation}
where
\begin{equation}
{\overline M}=M\biggl(1+\sqrt{\frac{M_0}{M}}\biggr).
\end{equation}
We obtain from Eq.(\ref{alphaeq}) and $k=2\times 10^{-13}$ for a
mean orbital cluster radius $\langle r_{\rm orb}\rangle_{\rm
cl}\sim 2\,{\rm Mpc}$:
\begin{equation}
\alpha_{\rm cl}\equiv (\sqrt{M_0})_{\rm cl}=3.1\times
10^{24}\,{\rm g}^{1/2}.
\end{equation}
For a cluster of mass $M_{\rm cl}\sim 10^{13}\,M_{\odot}$, we
obtain
\begin{equation}
\biggl(\sqrt{\frac{M_0}{M}}\biggr)_{\rm cl}\sim 22.
\end{equation}
We see that ${\overline M}\sim 22M$ and we can explain the
increase in the light bending without exotic dark matter.

From the formula Eq.(\ref{accelerationlaw2}) for $r\gg r_0\sim
14\,{\rm kpc}$ we get
\begin{equation}
a(r)=-\frac{G_0\overline M}{r^2}.
\end{equation}
We expect to obtain from this result a satisfactory description
of lensing phenomena using Eq.(\ref{lensingformula}).

The scaling by the parameter $\alpha_i=(\sqrt{M_0})_i$ caused by
the varying strength of the coupling of the skew field
$g_{[23]}(r)=f(r)\sin\theta$ to matter due to the renormalized
gravitational constant is seen to play an important role in
describing consistently the solar system and the galaxy and
cluster dynamics, without the postulate of exotic dark matter.

\section{Conclusions}

There is a large enough sample of galaxy data which fits our
predicted NGT acceleration law to warrant taking seriously the
proposal that NGT can explain the flat rotational velocity curves
of galaxies without exotic dark matter. We do predict that there
will be galaxy matter additional to that due to visible stars and
baryons, associated with the energy density $\rho_m$ residing in
the skew field $g_{[\mu\nu]}$. It is interesting to note that we
can fit the rotational velocity data of galaxies in the distance
range $0.02\,{\rm kpc} < r < 70\,{\rm kpc}$ and in the mass range
$10^5\, M_{\odot}< M < 10^{11}\,M_{\odot}$. without exotic dark
matter halos. We are required to investigate further the behavior
of the NGT predictions for distances approaching the cores of
galaxies, using a disk profile density. The lensing of clusters
can also be explained by the theory without exotic dark matter in
cluster halos.

We are able to obtain agreement with the observations in the solar
system and terrestrial gravitational experiments for suitable
values of the parameter $M_0$. This required that we scale $G$ and
the parameter $\sqrt{M_0}$ as functions of the mean orbital radius
$\langle r_{\rm orb}\rangle$ of bound systems with the behavior
$\alpha_i=(\sqrt{M_0})_i=\langle r_{\rm orb}\rangle_i^{3/2}$.

A numerical solution of the NGT field equations for cosmology must
be implemented to see whether the theory can account for the large
scale structure of the universe and account for galaxy formation
and big bang nucleosynthesis, without requiring the existence of
exotic, undetected dark matter and a positive cosmological
constant to describe dark energy.

\section{Appendix A: The Static Spherically Symmetric Solution
and the Vacuum Field Equations}

In the case of a spherically symmetric static
field~\cite{Moffat2}, the canonical form of $g_{\mu\nu}$ in NGT is
given by \begin{equation}
g_{\mu\nu}=\left(\matrix{-\alpha&0&0&w\cr
0&-\beta&f\hbox{sin}\theta&0\cr 0&-f\hbox{sin}\theta&
-\beta\hbox{sin}^2 \theta&0\cr-w&0&0&\gamma\cr}\right),
\end{equation} where $\alpha,\beta,\gamma$ and $w$ are functions
of $r$. The tensor $g^{\mu\nu}$ has the components:
\begin{equation} g^{\mu\nu}=\left(\matrix{{\gamma\over w^2-
\alpha\gamma}&0&0&{w\over w^2-\alpha\gamma}\cr 0&-{\beta\over
\beta^2+f^2}&{f\hbox{csc}\theta\over \beta^2+f^2}&0\cr
0&-{f\hbox{csc}\theta\over
\beta^2+f^2}&-{\beta\hbox{csc}^2\theta\over
\beta^2+f^2}&0\cr-{w\over w^2-\alpha\gamma}&0&0&-{\alpha\over
w^2-\alpha\gamma}\cr}\right). \end{equation} For the theory in
which there is no NGT magnetic monopole charge, we have $w=0$
and only the $g_{[23]}$ component of $g_{[\mu\nu]}$ survives.

The time independent field equations, Eq.(\ref{emptyspace}), in
empty space are given by
\begin{equation}
R_{11}(\Gamma)=-{1\over
2}A^{''}-{1\over 8}[(A^\prime)^2+4B^2] +{\alpha^\prime
A^\prime\over 4\alpha}+{\gamma^\prime\over
2\gamma}\biggl({\alpha^\prime\over 2\alpha} -{\gamma^\prime\over
2\gamma}\biggr) \end{equation} \begin{equation}
-\biggl({\gamma^\prime\over 2\gamma}\biggr)^\prime =-\Lambda
\alpha+{1\over 4}\mu^2{\alpha f^2\over \beta^2+f^2},
\end{equation} \begin{equation} {1\over
\beta}R_{22}(\Gamma)={1\over
\beta}R_{33}(\Gamma)\hbox{cosec}^2\theta={1\over\beta}+ {1\over
\beta}\biggl({2fB-\beta A^\prime\over 4\alpha}\biggr)^\prime
+{2fB-\beta A^\prime\over
8\alpha^2\beta\gamma}(\alpha^\prime\gamma +\gamma^\prime\alpha)
\end{equation} \begin{equation} +{B(fA^\prime+2\beta B)\over
4\alpha\beta} =-\Lambda-{1\over 4}\mu^2{f^2\over \beta^2+f^2},
\end{equation} \begin{equation} R_{00}(\Gamma)
=\biggl({\gamma^\prime\over 2\alpha}\biggr)^\prime
+{\gamma^\prime\over 2\alpha}\biggl({\alpha^\prime\over 2\alpha}
-{\gamma^\prime\over 2\gamma}+{1\over
2}A^\prime\biggr)=\Lambda\gamma -{1\over 4}\mu^2{\gamma f^2\over
\beta^2+f^2}, \end{equation} \begin{equation} R_{[10]}(\Gamma)=0,
\end{equation} \begin{equation} R_{(10)}(\Gamma)=0,
\end{equation}
\begin{equation}
\label{[23]equations}
R_{[23]}(\Gamma)=\hbox{sin}\theta\biggl[\biggl({fA^\prime+2\beta
B\over 4\alpha}\biggr)^\prime+{1\over 8\alpha}(fA^\prime+2\beta
B) \biggl({\alpha^\prime\over
\alpha}+{\gamma^\prime\over\gamma}\biggr) \end{equation}
\begin{equation} -{B\over 4\alpha}(2fB-\beta
A^\prime)=\biggl[\Lambda f -{1\over
4}\mu^2f\biggl(1+{\beta^2\over \beta^2+f^2}\biggr)\biggr]
\hbox{sin}\theta.
\end{equation}
Here, prime denotes
differentiation with respect to $r$, and we have used the
notation
\begin{equation} A=\hbox{ln}(\beta^2+f^2),
\end{equation}
\begin{equation} B={f\beta^\prime-\beta
f^\prime\over \beta^2+f^2}.
\end{equation}

Let us assume the long-range approximation for which the $\mu^2$ contributions
in the vacuum field equations (\ref{emptyspace}) can be
neglected and that $\mu^{-1} > 2M$. We then obtain the
static, spherically symmetric Wyman solution for
$\Lambda=0$~\cite{Wyman}: \begin{equation} \gamma=\exp(\nu),
\end{equation} \begin{equation}
\alpha=\frac{(f^2+\beta^2)(\gamma')^2}{4M_1^2\gamma},
\end{equation} \begin{equation}
f+i\beta=\frac{M_1^2(1+is)\exp(-\nu)}{A+i}{\rm
sech}^2[-\frac{1}{2}(1+is)^{1/2}\nu+B], \end{equation} where
$M_1,A,B$ and $s$ are integration constants and $\nu$ is an
arbitrary function of $r$.

We shall restrict our attention to the asymptotic condition
$g_{(\mu\nu)}\rightarrow\eta_{\mu\nu}$ where $\eta_{\mu\nu}$ is
the Minkowski flat metric tensor $\eta_{\mu\nu}={\rm
diag}(1,-1,-1,-1)$. This requires that
\begin{equation}
\sinh^2B=-1,
\end{equation}
so that we obtain
\begin{equation}
f+i\beta=-\frac{M_1^2(1+is)\exp(-\nu)}{A+i}{\rm csch}^2
[-\frac{1}{2}(1+is)^{1/2}\nu].
\end{equation} This form of
$f+i\beta$ does not place any {\it a priori} boundary condition on
$g_{[\mu\nu]}$.

We now obtain the form of the static Wyman solution:
\begin{equation}
\gamma=\hbox{exp}(\nu),
\end{equation}
\begin{equation}
\alpha=\frac{M_1^2(\nu^\prime)^2\hbox{exp}(-\nu)(1+s^2)}{A+i}
[\hbox{cosh}(a\nu)-\hbox{cos}(b\nu)]^{-2},
\end{equation}
\begin{equation}
f=\frac{2M_1^2}{1+A^2}\hbox{exp}(-\nu)[(1-As)\hbox{sinh}(a\nu)\hbox{sin}(b\nu)
$$ $$
+s(1-\hbox{cosh}(a\nu)
\hbox{cos}(b\nu)][\hbox{cosh}(a\nu)-\hbox{cos}(b\nu)]^{-2},
\end{equation}
where
\begin{equation}
a=\biggl({\sqrt{1+s^2}+1\over 2}\biggr)^{1/2},\quad
b=\biggl({\sqrt{1+s^2}-1\over 2}\biggr)^{1/2},
\end{equation}
and $\nu$ is implicitly determined by the equation
\begin{equation}
\hbox{exp}(\nu)(\hbox{cosh}(a\nu)-\hbox{cos}(b\nu))^2{r^2(1+A^2)\over
2M_1^2}=
(1-As)[\hbox{cosh}(a\nu)\hbox{cos}(b\nu)-1]
$$ $$
+(A+s)\hbox{sinh}(a\nu)\hbox{sin}(b\nu).
\end{equation}

At large distances $r\rightarrow\infty$, we have
\begin{equation}
\alpha=1+\frac{2M_1}{(1+A^2)^{1/2}}\frac{1}{r}+O\biggl(\frac{1}{r^2}\biggr),
\end{equation}
\begin{equation}
\gamma=1-\frac{2M_1}{(1+A^2)^{1/2}}\frac{1}{r}+O\biggl(\frac{1}{r^3}\biggr),
\end{equation}
\begin{equation}
f=-Ar^2+\frac{1}{3}sM_1^2+O\biggl(\frac{1}{r}\biggr).
\end{equation}
If we assume that $A=0$, then $M=M_1$ where $M$
denotes the mass of the particle source. \vskip 0.2 true in {\bf
Acknowledgments} \vskip 0.2 true in

This work was supported by the Natural Sciences and Engineering
Research Council of Canada. I thank Hilary Carteret for help with
the use of Maple 9 software and Gilles Esposito-Far\'ese, Gary
Mamon, Stacey McGough, Martin Green and Lee Smolin for helpful
discussions.

  \pagebreak \begin{center} Table 1. Values
of the total galaxy mass $M$ used to fit rotational velocity data. Also
shown are the mass-to-light-ratios $M/L$ with $L$ obtained from
ref.~\cite{McGaugh}. \vskip 0.3 true in \begin{tabular}{|l||c||c|}\hline
Galaxy & M($\times 10^{10}M_{\odot}$)& M/L($M_{\odot}/L_{\odot}$)\\\hline
NGC 2903 & 5.63 & 3.68\\ \hline NGC 5533 & 24.2 & 4.29\\ \hline NGC 5907 &
11.8 & 4.92\\ \hline NGC 6503 & 1.36 & 2.84\\ \hline NGC 3198 & 3.0 &
3.33\\ \hline NGC 2403 & 2.37 & 3.0\\ \hline NGC 1560 & 0.427 & 12.20\\
\hline NGC 4138 & 2.94 & 3.59\\ \hline NGC 3379 & 5.78 & --\\ \hline M33 & 0.93 & 1.98\\
\hline UGC 6917 & 0.96 & 2.53\\ \hline UGC 6923 & 0.388 & 1.76\\
\hline UGC 6930 & 1.04  & 2.08\\ \hline FORNAX & 0.0026 & 1.86\\ \hline DRACO & 0.00050 & 27.94\\
\hline $\omega$ Centauri & $3.05\times 10^{-5}$ & -- \\\hline
\end{tabular}
\end{center}

\pagebreak
\begin{center}
\includegraphics[width=2.5in,height=2.5in]{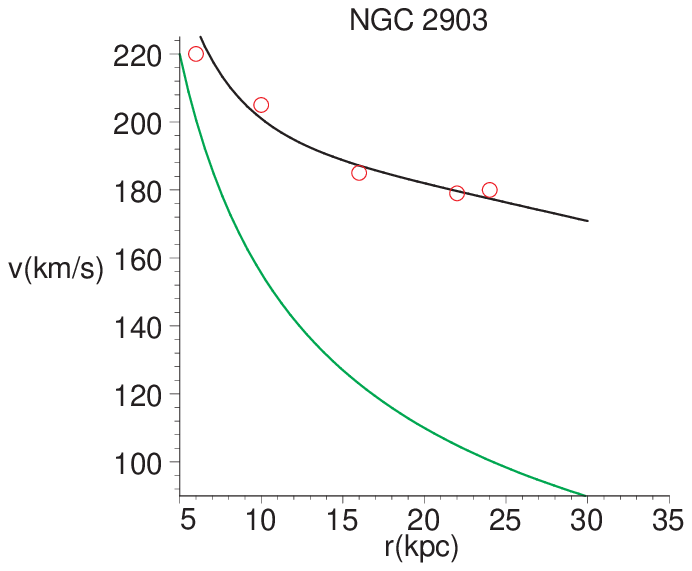}
\includegraphics[width=2.5in,height=2.5in]{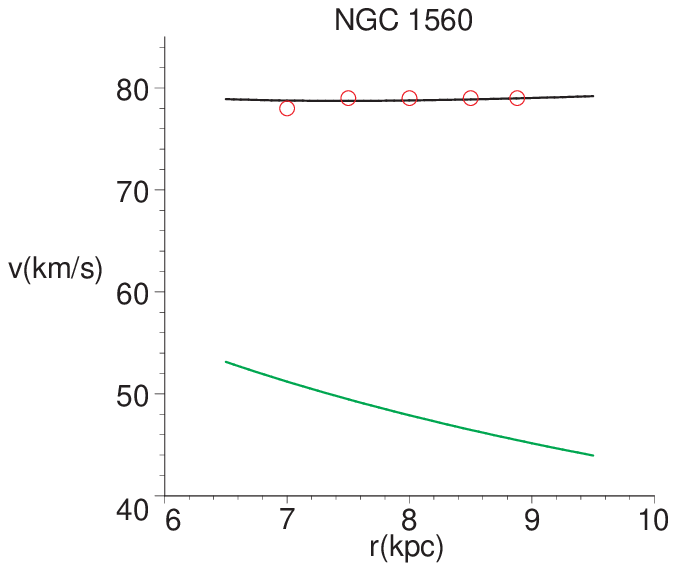}
\end{center}
\vskip 0.1 in
\begin{center}
\includegraphics[width=2.5in,height=2.5in]{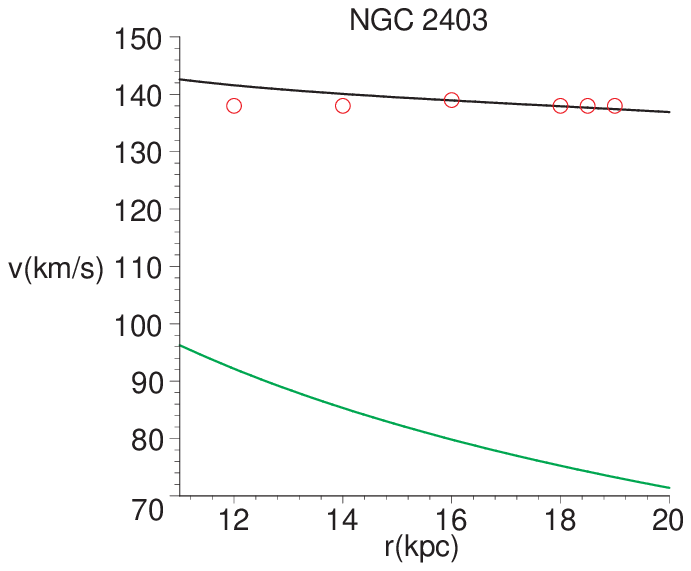}
\includegraphics[width=2.5in,height=2.5in]{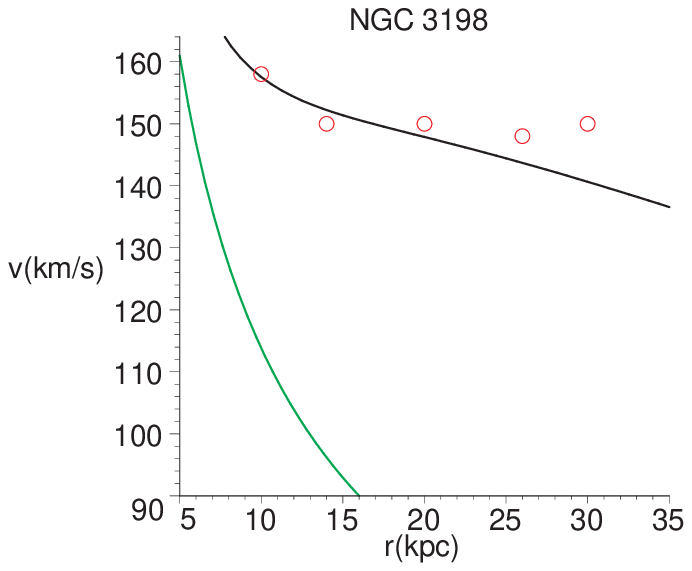}
\end{center}
\vskip 0.1 true in
\begin{center}
\includegraphics[width=2.5in,height=2.5in]{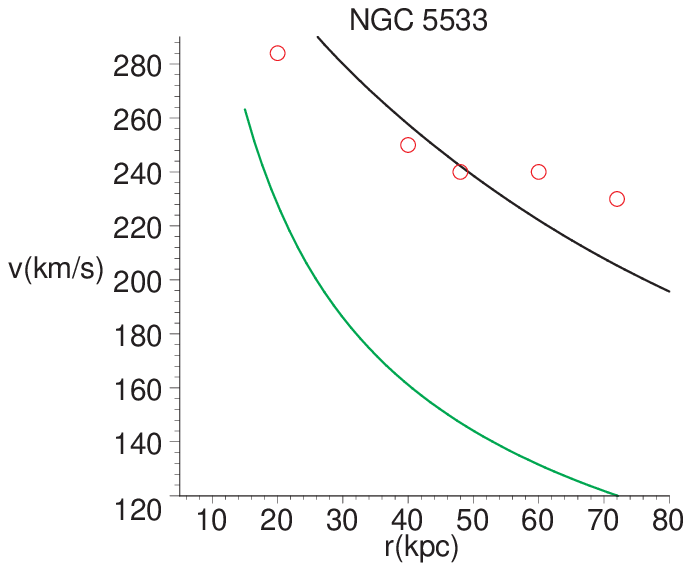}
\includegraphics[width=2.5in,height=2.5in]{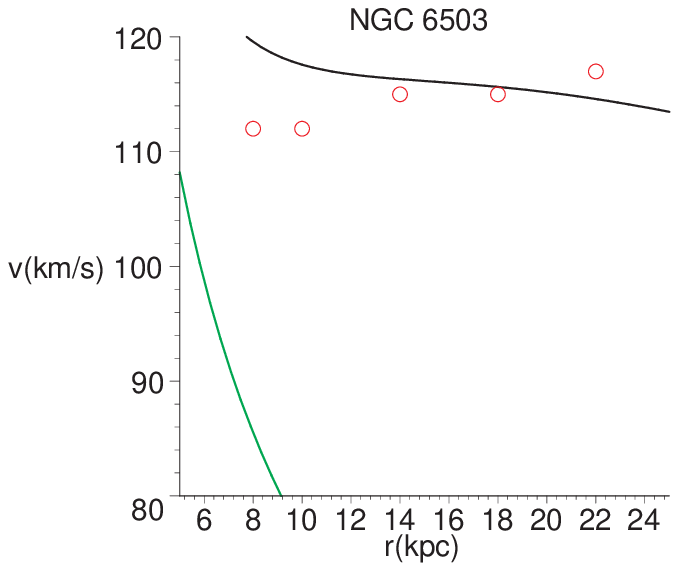}
\end{center}
\vskip 0.1 true in
\begin{center}
\includegraphics[width=2.5in,height=2.5in]{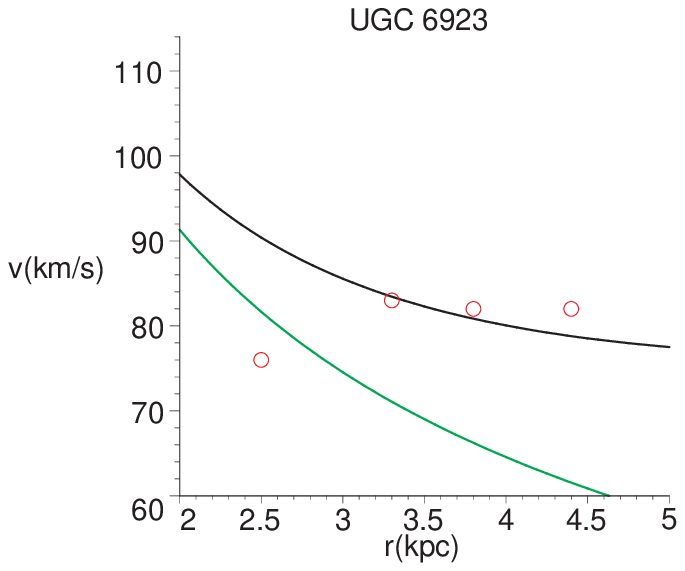}
\includegraphics[width=2.5in,height=2.5in]{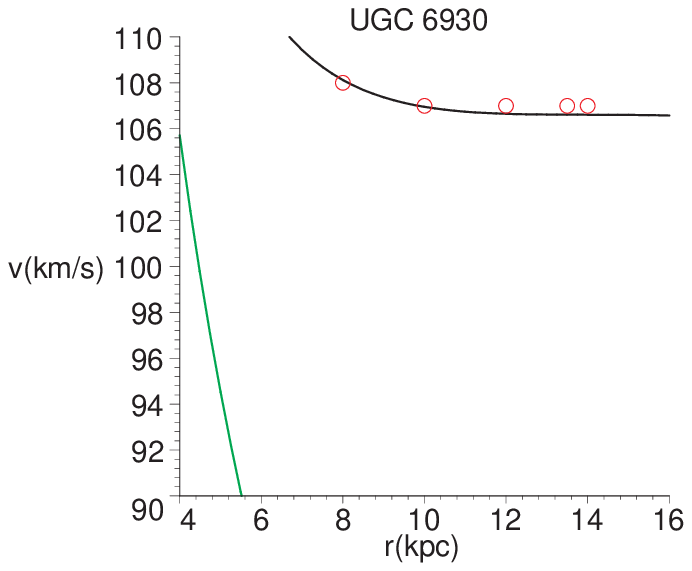}
\end{center}
\vskip 0.1 true in
\begin{center}
\includegraphics[width=2.5in,height=2.5in]{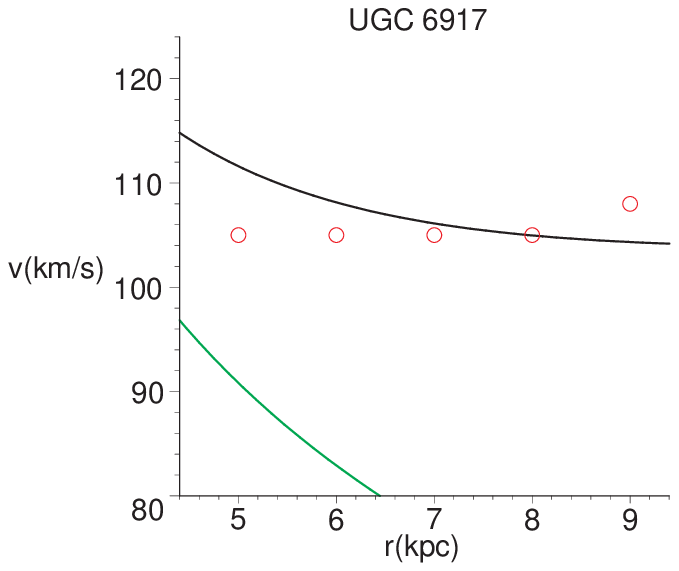}
\includegraphics[width=2.5in,height=2.5in]{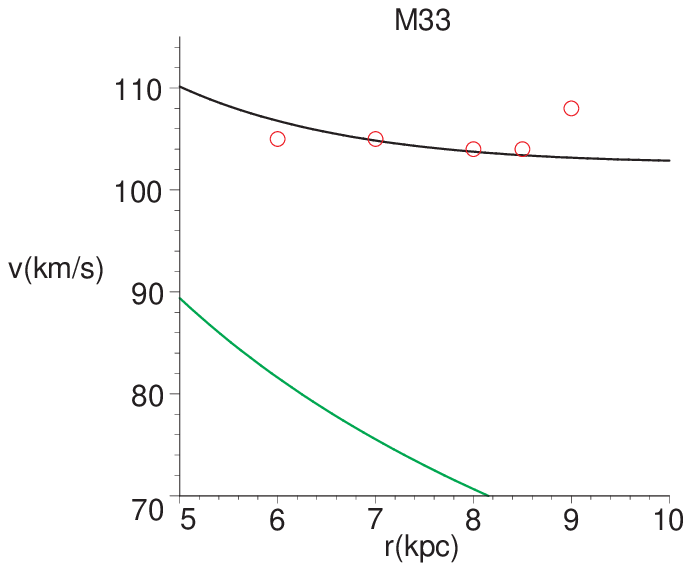}
\end{center}
\vskip 0.1 true in
\begin{center}
\includegraphics[width=2.5in,height=2.5in]{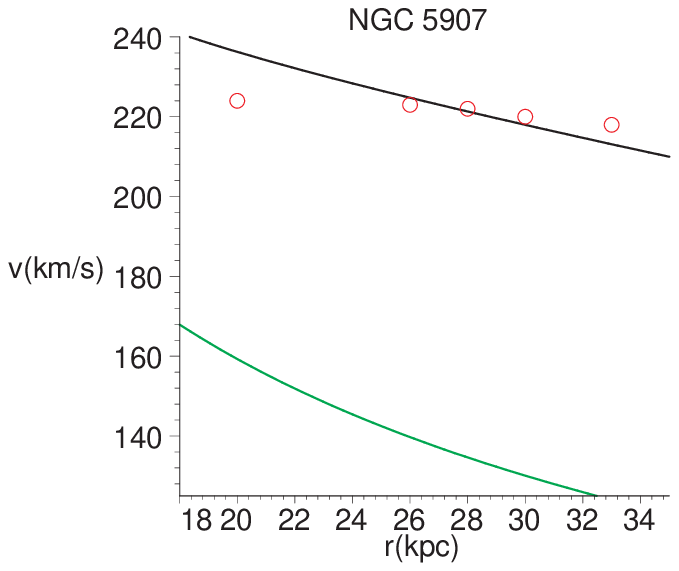}
\includegraphics[width=2.5in,height=2.5in]{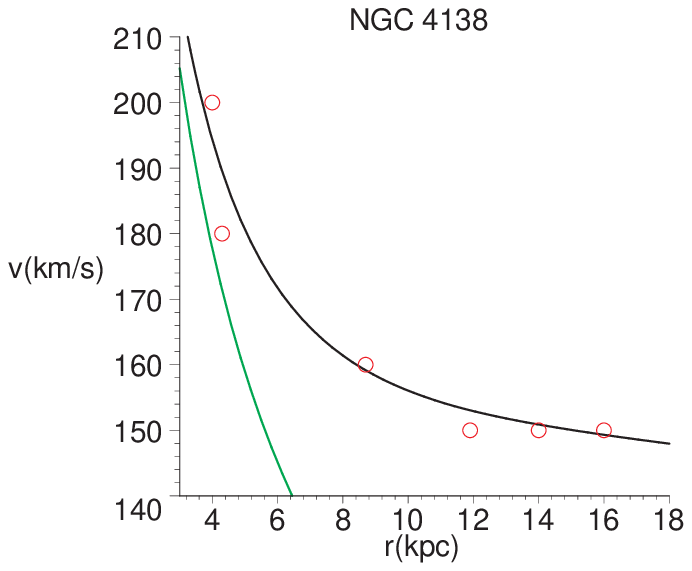}
\end{center}
\vskip 0.1 true in
\begin{center}
Fig. 1 - Fits to low-surface-brightness and
high-surface-brightness spiral galaxy data. The black curve is the
rotational velocity $v$ versus $r$ obtained from the modified
Newtonian acceleration, while the green curve shows the Newtonian
rotational velocity $v$ versus $r$. The data are shown as red
circles.
\end{center}
\vskip 0.1 true in
\begin{center}
\includegraphics[width=2.5in,height=2.5in]{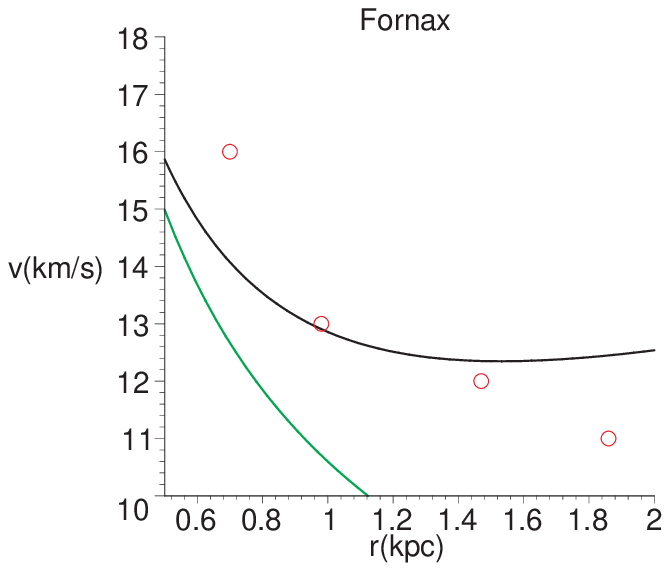}
\includegraphics[width=2.5in,height=2.5in]{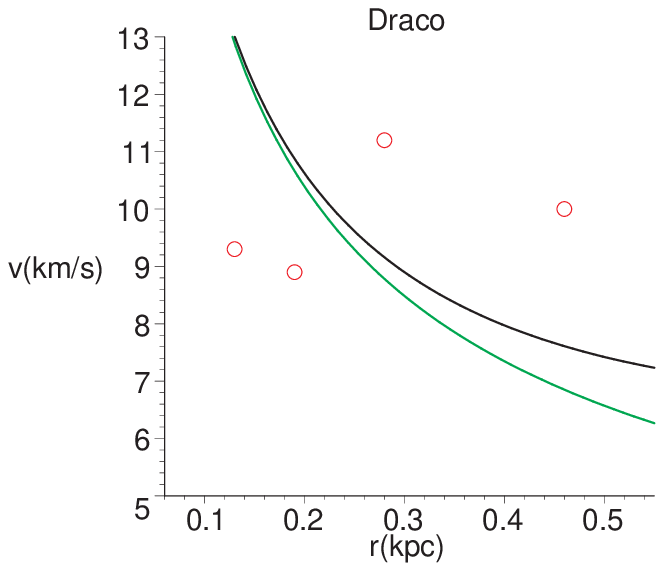}
\end{center}
\vskip 0.1 true in
\begin{center}
Fig. 2 - Fits to the data of two dwarf galaxies Fornax and Draco.
The simple relation $V\sim \sqrt{2}\sigma$ is assumed between the
velocity dispersion $\sigma$ and the rotational velocity $v$. The
black curve is the rotational velocity $v$ versus $r$ obtained
from the modified Newtonian acceleration, while the green curve
shows the Newtonian rotational velocity $v$ versus $r$. The data
are shown as red circles and the errors (not shown) are large and
for Draco the Newtonian fit cannot be distinguished from the NGT
fit to the data within the errors.
\end{center} \vskip 0.1 true in
\begin{center}
\end{center}
\vskip 0.1 true in
\vskip 0.1 true in
\begin{center}
\includegraphics[width=2.5in,height=2.5in]{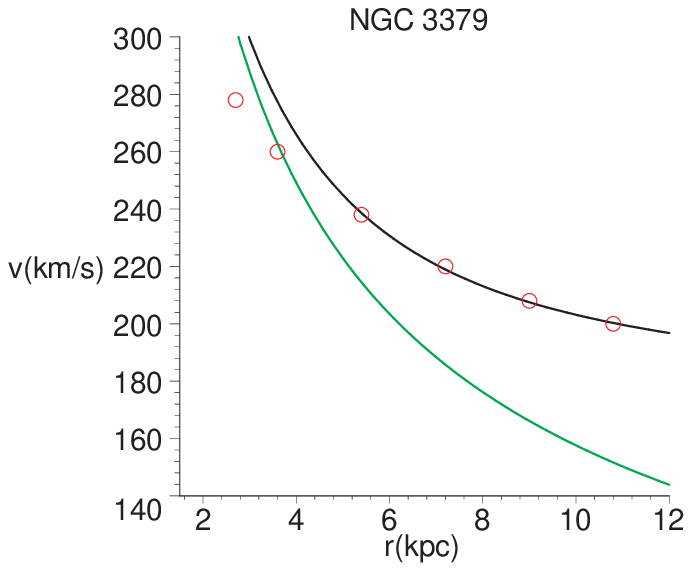}
\includegraphics[width=2.5in,height=2.5in]{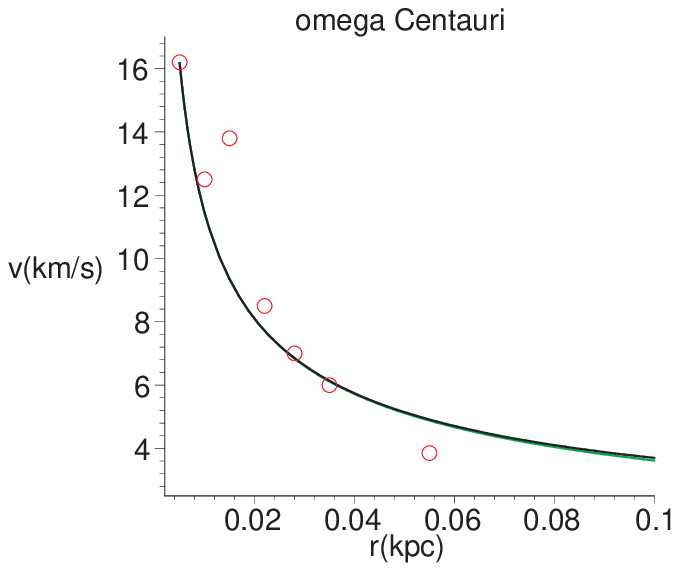}
\end{center}
\vskip 0.1 true in
\begin{center}
Fig. 3 - Fits to the elliptical galaxy NGC 3379 and the globular
cluster $\omega$ Centauri. The black curve is the rotational
velocity $v$ versus $r$ obtained from the modified NGT
acceleration, while the green curve is the Newtonian-Kepler
velocity curve. For $\omega$ Centauri the black and green curves
cannot be distinguished from one another.
\end{center}
\vskip 0.1 true in
\begin{center}
\includegraphics[width=2.5in,height=2.5in]{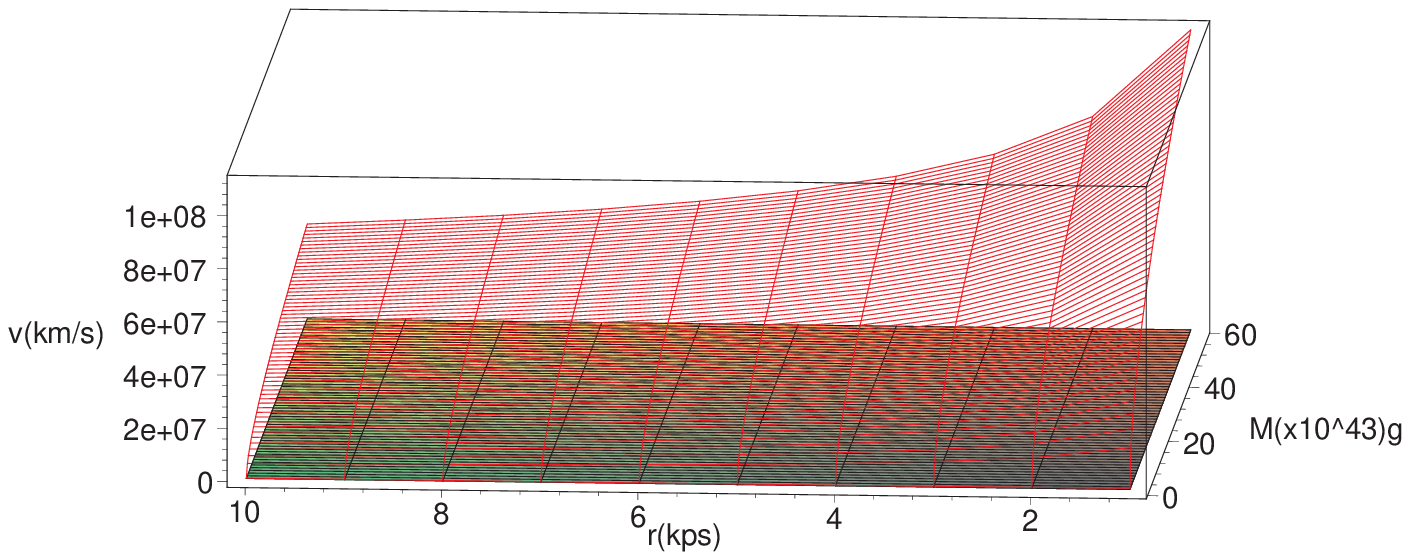}
\end{center}
\vskip 0.1 true in
\begin{center} Fig. 4 -
3-dimensional plot of $v$ versus the range of distance $0.1\,{\rm
kpc}<r<10\,{\rm kpc}$ and the range of galaxy mass $5\times
10^6\,M_{\odot} < M < 2.5\times 10^{11}\,M_{\odot}$. The red
surface shows the Newtonian values of the rotational velocity $v$,
while the black surface displays the NGT prediction for $v$.
\end{center}

\end{document}